\documentclass[preprint,preprintnumbers,amsmath,amssymb,prd]{revtex4}
\usepackage{amsmath,mathrsfs,amsbsy,color,graphicx,bm,amsthm,amsfonts}
\usepackage{amsfonts}
\usepackage{amssymb}
\usepackage{graphicx}
\usepackage{subfigure}
\usepackage{amsmath,amssymb,epsfig}
\usepackage{color}
\usepackage{ulem}

\newcommand{\be}{\begin{equation}}
\newcommand{\ee}{\end{equation}}
\newcommand{\bea}{\begin{eqnarray}}
\newcommand{\eea}{\end{eqnarray}}
\newcommand{\ba}{\begin{array}}
\newcommand{\ea}{\end{array}}

\usepackage{epstopdf}
\usepackage{hyperref}
\begin{document}

\title{Optimal quantum strategy for locating Unruh channels}

\author{Qianqian Liu$^{1}$, Tonghua Liu$^{2}$\footnote{Email: liutongh@yangtzeu.edu.cn}, Cuihong Wen$^{1}$, and Jieci Wang$^{1}$\footnote{Email: jcwang@hunnu.edu.cn}}
\affiliation{$^1$  Department of Physics and Synergetic Innovation Center for Quantum Effects,\\Key Laboratory of Low-Dimensional Quantum Structures and Quantum Control of Ministry of Education, \\
Key Laboratory for Matter Microstructure and Function of Hunan Province,\\
 Hunan Normal University, Changsha 410081, China}
\affiliation{$^2$ School of Physics and Optoelectronic,
Yangtze University, Jingzhou 434023, China
	}

\begin{abstract}
From the perspective of quantum information theory, the effect of Unruh radiation on a two-level accelerated detector can be modeled as a quantum channel. In this work, we employ the tools of channel-position finding  to locate Unruh channels. The signal-idler  and idler-free protocols are explored to determine the position of the target Unruh channel within a sequence of background channels. We derive the fidelity-based bounds for the ultimate error probability of each strategy and obtain the conditions where the signal-idler protocol is superior to the protocol involving idler-free states. It is found that the lower bound of the error probability for the signal-idler scheme exhibits clear advantages in all cases, while the idler-free scheme can only be implemented when the temperature of the two channels is very close and the number of initial states is insufficient. Interestingly, it is shown that the optimal detection protocol relies on the residual correlations shared between the emitted probe state and the retained idler modes.

\end{abstract}
\vspace*{0.5cm}

\maketitle
\section{Introduction}\label{section1}

%This series of theory studies on the Unruh effect is essential for understanding more complex phenomena such as black hole dynamics and the quantum nature of relativistic gravity.
The Unruh effect \cite{Unruh1,UD1,HKG,Unruh2} is one of the most monumental achievements of quantum field theory in curved spacetime. It plays a crucial role in the understanding of vacuum fluctuations and the nature of quantum thermal effects. It was predicted that a uniformly accelerated observer will detect a thermal bath from expressing the vacuum state in terms of a different set of operator basis defined along the time-like killing vector in their locally accelerated coordinate system \cite{gendeg1, gendeg2, exp3,curved1,curved2}. A variety of techniques have been employed to analyze this phenomenon including the response of a two level system, referred to as an Unruh-DeWitt (UD) detector \cite{UD2, UD3, UD4, coupling3}, when it absorbs these thermal particles.  Studying the Unruh effect from the perspective of quantum information theory could not only be helpful in understanding the Hawking effect  \cite{HK1,HK2,MK}, but also provides an explanation for the generation and degradation of entanglement in curved spacetime \cite{curved4,curved5,curved6}.
 Direct observation of the Unruh effect is considered as one of the key experimental goals of contemporary physics  \cite{experimental1,experimental3,experimental4,experimental2,open1}. However, a simple calculation shows that a 1 Kelvin Unruh temperature corresponds to an acceleration of the order of $\sim 10^{21}m/s^{2}$, which is extremely challenging to obtain \cite{experimental4,open1}. In this sense, the technical obstacles to the detection of Unruh radiation lead to the indistinguishability of Unruh channel in general relativistic background.

\iffalse On the other hand, the quantum channel discrimination aims to discriminate quantum channels \cite{CPF0,CPF1,CPF2,CPF3,channel1,channel2}.
The channel position finding (CPF) aims to discriminate different physical processes, modeled as quantum channels \cite{channel1,channel2}, from different physical processes \cite{CPF1,CPF2,CPF3}.  in which a sequence of channels is given that all of which are identical except one, and the task is to locate the unique channel.\fi

On the other hand, quantum channels can model various physical processes \cite{channel1,channel2}, so the discrimination of different quantum channels \cite{CPF0,CPF1,CPF2,CPF3} is a fundamental task in quantum information theory. The theory of channel-position finding (CPF) has been effectively used to determine the target channel with varying transmittance or induced noise within a range of background loss channels \cite{CPF1,CPF2,CPF3}. Recently, the advantages of quantum entanglement have been demonstrated in CPF, for example the thermal loss channel \cite{channel1} and the amplitude damping channel \cite{DC,ADC}. In this paper, we study the task of determining the location of Unruh channels, in which different  accelerations would  induce differentiated responses in the detectors \cite{noisy1, UD2,UD3,coupling3}. In the UD detector model, the divergence in temperature predictions among detectors with different acceleration simplifies the task from identifying the  channel temperatures to identifying the detector's accelerations. \iffalse The motivation for our research is to find the optimal strategy for locating the position of a target Unruh channel which is hidden among  different thermal background channels.\fi

Here we focus on the problem of CPF under the constraint that the sources considered are comprised of at most one photon. Two protocols will be considered: the signal-idler (SI) protocol and idler-free (IF) protocol. In fact, in the applications of quantum sensing, the assistance of idler modes has been a crucial feature to achieve quantum enhanced performance \cite{SI1,QI1}, but the IF channel identification schemes have also received a lot of attention because of their ability to eliminate quantum memory \cite{CPF2,IF1}.
We consider two scenarios: (i) The temperature of the target channel is zero, and it is located within a series of reference channels; (ii) the temperature difference between the target channel and the reference channel is particularly small. These two scenarios effectively encompass the potential background in which the target channel may exist.
We establish fidelity-based bounds on the final error probability in the multiple channel discrimination problem and identify the quantum dominance involving various quantum sources.
The main purpose of our study is to find the optimal strategy for locating the Unruh channels, and the optimal operating conditions for different strategies. \iffalse
Our analysis demonstrates that the lower bound of the error probability of the SI protocol has obvious advantages in all cases, while the IF protocol is only suitable for resolving the situation when the temperature of the two channels is very close and the number of initial states is insufficient. Most importantly, the performance is closely related with the residual correlation of the system.\fi

This paper is organized as follows: Sec. II outlines the model of Unruh channel location and calculates the detection error probabilities. In Sec. III, we compare the advantage of detection error probabilities between the SI protocol and the IF protocol, and Sec. IV presents the conclusions. Throughout the paper, we adopt the conventions $\hbar=G=c=\kappa_{B}=1$.

\section{The model of Unruh channel location}\label{sec:probspec}
\subsection{The scheme}
\begin{figure}[ht]
\centering
\includegraphics[width=0.52\textwidth]{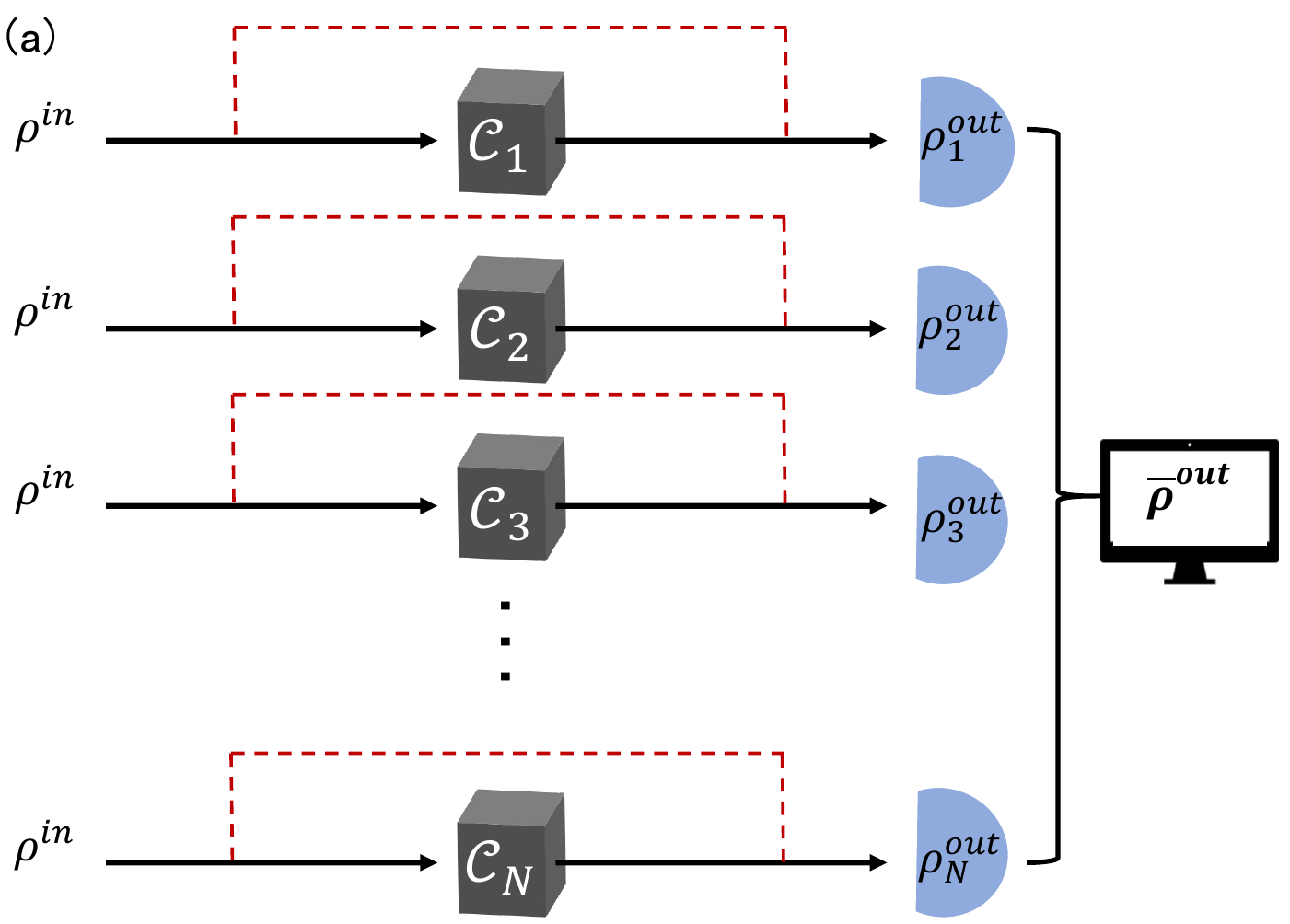}
\includegraphics[width=0.41\textwidth]{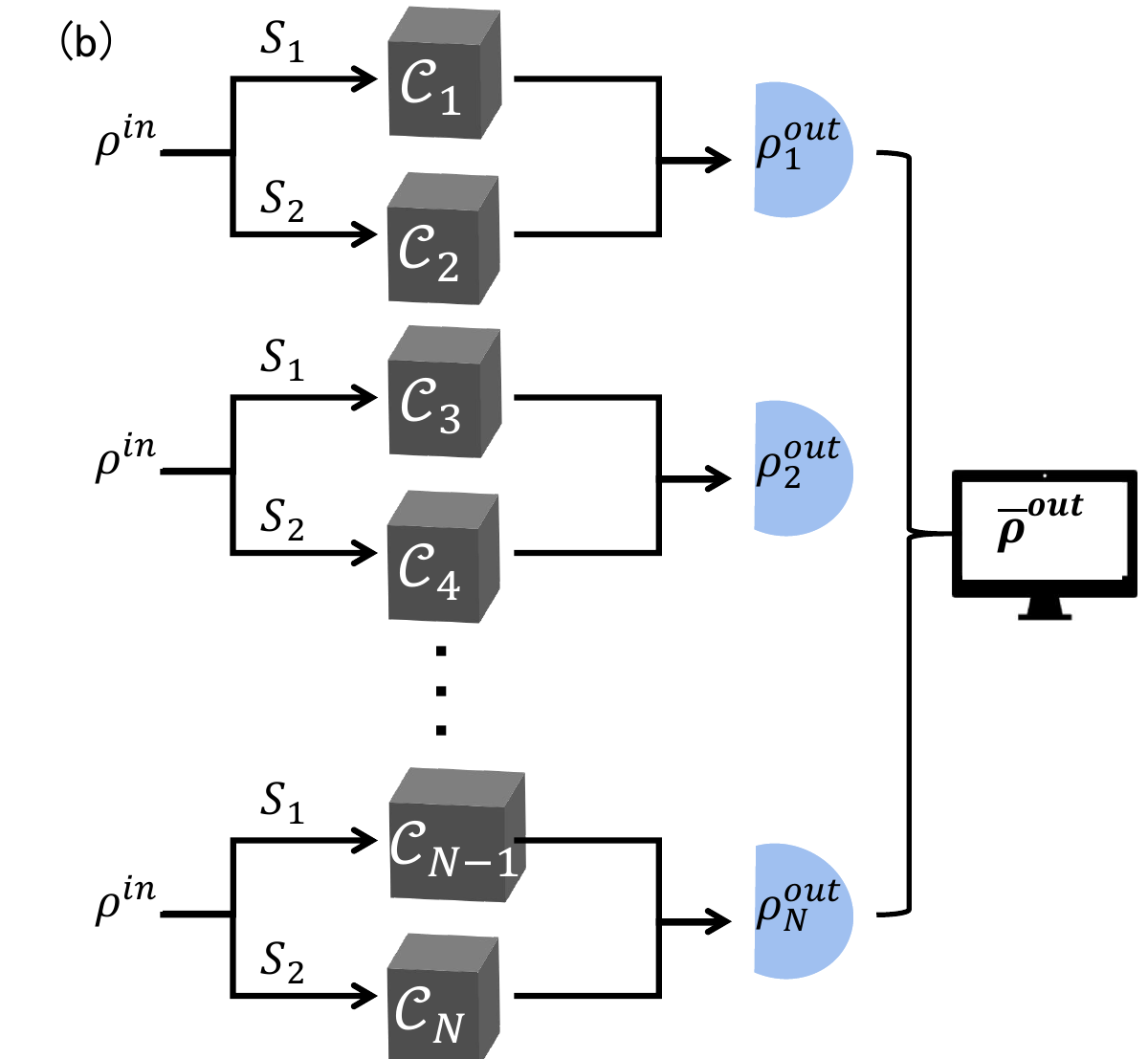}
\caption{ Our setup for the CPF protocol that provides a benchmark for the general quantum channel. We assume $N$ boxes, consisting of a target channel and $N-1$ reference channels.
Two different protocols for the CPF of Unruh channel are employed. In panel (a), we utilize the signal-idler protocol of CPF, while in panel (b), we employ the idler-free protocol of CPF for biphoton states.
}\label{fig1}
\end{figure}

In this paper, we employ the tools of CPF involving $N\geq2$ boxes to locate the target Unruh channel position. As shown in Fig. (\ref{fig1}), the boxes $\mathcal{C}_i(i=1,2,...,N)$
are modeled as Unruh channels with different temperatures. The characterization of each channel is presented in the Appendix A.  The target channel $\mathcal{C}_\mathcal{T}$ occupies one box with acceleration $q_{i}$, while the other $N-1$ boxes represent the reference channel $\mathcal{C}_\mathcal{R}$ with acceleration $q_{j\neq i}$.
Identification of the target Unruh channel is a symmetric hypothesis testing problem where the task is to discriminate between $N$ hypotheses given by \cite{symme1,symme2}
\begin{equation}\label{GLB}
H_{i}:\mathcal{C}_{i}=\mathcal{C}_\mathcal{T},\mathcal{C}_{j\neq i}=\mathcal{C}_\mathcal{R}.
\end{equation}
At the transmitter, the initial state $\rho^{in}$ injects into each of the boxes.
 Each channel is represented by an accelerated detector which interacts with its surroundings \cite{coupling3}, and this detection process can be described as a quantum map for the quantum state, as detailed in the Appendix A.
The task of correctly identifying the target channel in a series of reference channels may be reduced to distinguishing the possible channel outputs $\mathcal{C}_\mathcal{T}(\rho)$ and $\mathcal{C}_\mathcal{R}(\rho)$ \cite{CPF1,CPF2,CPF3}.
Suppose that the overall input state has a tensor product form over the $N$ boxes, such that $\rho=\sigma^{\bigotimes N}$, and utilizing $M\gg1$ identical transmissions of these types.
We assume equiprobable hypotheses $p_{i}=N^{-1}$ for any $i$, and then compute the error probabilities $p_{\mathrm{err}}^{N,M} (\rho)$.
 Obtaining an exact analytical bound of the error probability is challenging.
However, the upper and lower bounds can be derived using the pretty good measurement (PGM) \cite{LUB1,symme2}
\begin{equation}\label{GLB}
p_{\mathrm{err}}^{N,M} (\rho) \leq (N-1) F^{2M}\left(\mathcal{C}_\mathcal{T}(\rho), \mathcal{C}_\mathcal{R}(\rho)\right),
\end{equation}
\begin{equation}\label{GUB}
p_{\mathrm{err}}^{N,M} (\rho) \geq \frac{N-1}{2 N} F^{4M}\left(\mathcal{C}_\mathcal{T}(\rho), \mathcal{C}_\mathcal{R}(\rho)\right),
\end{equation}
where $F(\rho, \sigma)$ is the Bures fidelity \cite{fidelity1,fidelity2}
\begin{equation}
F(\rho, \sigma):=\|\sqrt{\rho} \sqrt{\sigma}\|_{1}=\operatorname{tr} \sqrt{\sqrt{\rho} \sigma \sqrt{\rho}}.
\end{equation}

\subsection{ The signal-idler protocol}\label{SECbipartite}

We first employ the SI strategy to locate the Unruh channels. Such protocol  has been proved to be effective in discriminating Gaussian lossy channels \cite{channel1,CPF3,ADC,DC,channel2}.
 As depicted in Fig. (\ref{fig1}a), the initial state of the entire system is prepared in tensor product over all the boxes ($\bigotimes N$), where each signal $S_{i}$ (black) box is entangled with an ancillary idler $I_{i}$ (red).
 The input state for each box is a maximally entangled state
\begin{eqnarray}\label{rhogen}
|\Psi^{in}\rangle
=\frac{1}{\sqrt{2}}(|00\rangle+|11\rangle).
		\end{eqnarray}
The SI strategy involves only the signal probing the box, while the idler state is directly sent to the receiver for combination with the output. The associated quantum channel is expressed as
\begin{equation}
\mathcal{E}_i^{N}:= \otimes_{j \neq i} (\mathcal{C_{R}}_{j} \otimes \mathcal{I}_{I_j}) \otimes (\mathcal{C_{T}}_{i} \otimes \mathcal{I}_{I_i}).
\label{bipartitechannel}
\end{equation}
Upon the action of an Unruh channel only on the signal (S) mode while performing the identity on the reference idler (I) mode, we obtain the density operator of the output state
\begin{equation}\label{rhogen'} \rho^{out}
=\frac{1-q}{2}(|00\rangle\langle00|+|00\rangle\langle11|+|11\rangle\langle00|+|11\rangle\langle11|)
+\frac{v^{2}}{2}|01\rangle\langle01|+\frac{q v^{2}}{2}|10\rangle\langle10|.
		\end{equation}
The fidelity between the two outputs under the Unruh channel with acceleration parameters $q_{1}$ and $q_{2}$ is found to be
\begin{equation}
\begin{aligned}\label{plottedfid}
F(\rho^{out}_{q_{1}},\rho^{out}_{q_{2}}) & = \frac{1}{4}\left[4(-1+q_{1})(-1+q_{2})+
(1+q_{1}q_{2})\nu^{4} \right]\\
& +\frac{1}{4}\left[\sqrt{-4(-2+q_{1})q_{1}-(1+q^{2}_{1})\nu^{4}}
\sqrt{-4(-2+q_{2})q_{2}-(1+q^{2}_{2})\nu^{4}}\right].
\end{aligned}
\end{equation}
%where $q_{1}$ and $q_{2}$ represent the acceleration parameters of the reference and target channels, respectively.
By inserting Eq. (\ref{plottedfid}) into Eqs. (\ref{GLB})-(\ref{GUB}), the error probability for the SI protocol
is then lower and upper bounded by
\begin{equation}
     p_{\mathrm{err}}^{N,M}(\rho) \geq \frac{N-1}{2N} F(\rho^{out}_{q_{1}},\rho^{out}_{q_{2}})^{4M},
\end{equation}
and
\begin{equation}
    p_{\mathrm{err}}^{N,M}(\rho) \leq (N-1) F(\rho^{out}_{q_{1}},\rho^{out}_{q_{2}})^{2M},
\end{equation}
respectively.

\subsection{ The idler-free protocol}\label{idler-free}

Then we consider the idler-free protocol without idler mode reservation. In this case, the two-mode state  $\rho^{\mathrm{in}}$ serves as probes into two adjacent boxes, with the modes labeled as signals
 $S_{1}$ and $S_{2}$, as illustrated in Fig. (\ref{fig1}b).
While the optimal quantum strategy for various scenarios, such as quantum illumination \cite{QI1,QI2,QI3}, spectroscopy \cite{Spectroscopy1}, and quantum readings \cite{QD1}, often involves an entangled idler-assisted protocol, the storage of the idler mode poses a challenging task. The character of IF protocols lies in their ease of implementation and the absence of considerations for memory construction. Therefore, investigating the IF strategy is valuable, as it will help us understand whether quantum superiority can still be achieved even without a quantum memory for storing the idler mode.

 In the IF protocol setup, we take advantage of the complete entanglement exhibited with the Bell state in the multichannel array to achieve quantum advantage.
 For any CPF problem consisting of $N\geq4$ (even) independent channels, the global quantum channel acting on the initial state is:
\begin{equation}
\mathcal{E}_i^{N/2}:= \otimes_{j \neq i} (\mathcal{C_{R}}_{j} \otimes \mathcal{C_{R}}_{j}) \otimes (\mathcal{C_{T}}_{i} \otimes \mathcal{C_{R}}_{i}).
\label{idlerfreechannel}
\end{equation}

%\begin{equation}
 %\mathcal{B}(\rho) = K_{1} \rho K_{1}^{\dag}+K_{2} \rho K_{2}^{\dag}+K_{1} \rho K_{2}^{\dag}+K_{2} \rho K_{1}^{\dag}+K_{3} \rho K_{3}^{\dag},
%\end{equation}
If both modes $S_{1}$ and $S_{2}$ pass through two Unruh channels with the same acceleration parameter $q_{1}$, the final state at the output is
\begin{equation}\label{IF(q1,q1)} \rho^{out}_{q_{1},q_{1}}=\frac{1}{2}
\left(	\begin{array}{cccc}
Q_{1}^{2}+\nu^{4} &  0 &  0 & Q_{1}^{2} \\
		0&  Q_{1} \nu^{2}+Q_{1}q_{1} \nu^{2}  & 0 & 0 \\
		0& 0 &  Q_{1} \nu^{2}+Q_{1}q_{1} \nu^{2} & 0 \\
Q_{1}^{2} &  0&  0& Q_{1}^{2}+q_{1}^{2}\nu^{4}
	\end{array}\right),
		\end{equation}
where $Q_{1}=1-q_{1}$.
If the two modes of the initial state $\rho^{\mathrm{in}}$ pass through two Unruh channels with acceleration $q_{1}$ and $q_{2}$, the final state at the output takes the form:
\begin{equation}\label{IF(q1,q2)} \rho^{out}_{q_{2},q_{1}}=\frac{1}{2}
\left(	\begin{array}{cccc}
Q_{1}Q_{2}+\nu^{4} &  0 &  0 & Q_{1}Q_{2} \\
		0&  Q_{1} \nu^{2}+Q_{2}q_{1} \nu^{2}  & 0 & 0 \\
		0& 0 &  Q_{2} \nu^{2}+Q_{1}q_{2} \nu^{2} & 0 \\
Q_{1}Q_{2} &  0&  0& Q_{1}Q_{2}+q_{1}q_{2}\nu^{4}
	\end{array}\right),
		\end{equation}
where $Q_{2}=1-q_{2}$.
The lower and upper bounds for the error probabilities of the correct channel pair, calculated based on the fidelity between the two output states mentioned, are as follows \cite{CPF3}:
\begin{equation}\label{idlerfreeerr1}
     \tilde{p}_{\mathrm{err}}^{N,M}(\rho) \geq \frac{N-2}{2N} F(\rho^{out}_{q_{1},q_{1}},\rho^{out}_{q_{2},q_{1}})^{4M},
\end{equation}
and
\begin{equation}\label{idlerfreeerr2}
    \tilde{p}_{\mathrm{err}}^{N,M}(\rho) \leq \frac{N-2}{2} F(\rho^{out}_{q_{1},q_{1}},\rho^{out}_{q_{2},q_{1}})^{2M}.
\end{equation}

The objective of the CPF task is to determine the location of the target channel, rather than merely identifying the pair containing it.
In an IF protocol, the first stage involves successfully identifying the correct pair, and the second stage entails engineering a secondary CPF protocol by combining the correct pair with two additional reference channels. This enables us to pinpoint the location of the target channel. To maintain the energy constraint, we choose to divide the total number of probes into two parts, generating $M/2$ probes for each stage of the IF strategy in the CPF process.

Considering this two-stage approach, there are two ways in which an overall error can occur. The first
scenario involves misidentifying the pair where the target channel is located in the initial stage, resulting in a failure to accomplish the task of locating the target channel. The second scenario involves correctly identifying the pair in which the target channel is located in the first stage, but in the second stage, there is an incorrect identification of which of the two channels is the target channel. Therefore, we can utilize the relevant lower and upper bounds to derive the final error probability of the IF scheme as follows:
\begin{equation}
\begin{split}
    p_{\mathrm{err}}^{N,M/2}&(|{\psi_2}\rangle\langle{\psi_2}|) = \tilde{p}_{\mathrm{err}}^{N,M/2}(|{\psi_2}\rangle\langle{\psi_2}|) + \left[1- \tilde{p}_{\mathrm{err}}^{N,M/2}(|{\psi_2}\rangle\langle{\psi_2}|) \right] \tilde{p}_{\mathrm{err}}^{4,M/2}(|{\psi_2}\rangle\langle{\psi_2}|) .
    \end{split}
\end{equation}

\section{Quantum advantage of the signal-idler strategy }\label{Quantum advantage}

In the previous section, we utilized the two-energy level detector as a thermometer model and the theory of CPF for discriminating the Unruh temperature.
 As mentioned earlier, to find the most effective method for detecting the Unruh channel, we need to compare the upper and lower bounds of error probabilities associated with different protocols.
 Denoting the upper and lower bounds of the SI (IF) protocols as $p_{\mathrm{err}}^{\mathrm{SI,U}}$ and $p_{\mathrm{err}}^{\mathrm{SI,L}}$ ($p_{\mathrm{err}}^{\mathrm{IF,U}}$ and $p_{\mathrm{err}}^{\mathrm{IF,L}}$), we define the minimum guaranteed advantage (MGA) as the minimum performance enhancement achieved by a SI strategy over the IF one \cite{fidelity2},
 \begin{equation}\label{MGA}
\Delta p_{\mathrm{err}}^{min}:=
p_{\mathrm{err}}^{\mathrm{IF,L}}-p_{\mathrm{err}}^{\mathrm{SI,U}}.
\end{equation}
If $\Delta p_{\mathrm{err}}^{min}>0$, the advantage of  SI strategy is guaranteed. One can also define the maximum potential advantage (MPA) as follows:
  \begin{equation}\label{MPA}
\Delta p_{\mathrm{err}}^{max}:=
p_{\mathrm{err}}^{\mathrm{IF,L}}-p_{\mathrm{err}}^{\mathrm{SI,L}}.
\end{equation}
 This represents the maximum potential improvement that quantum strategies can bring when the derived lower bound is fundamental.

 \begin{figure}[ht]
\centering
\includegraphics[width=0.49\textwidth]{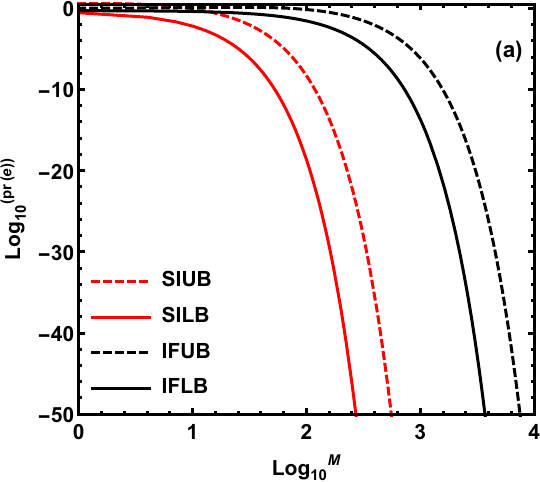}
\includegraphics[width=0.49\textwidth]{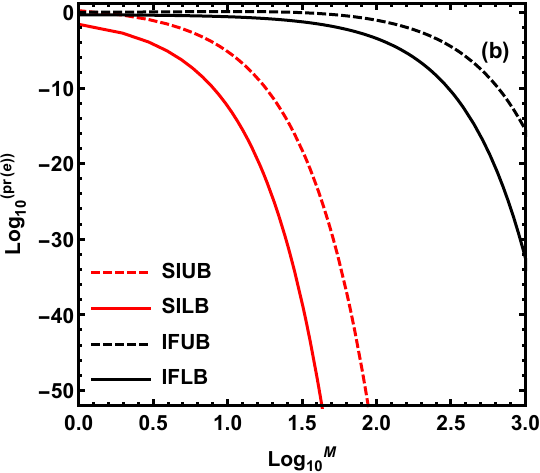}
\caption{(Color online) Quantum channel-position finding error probability $P^{N,M}_{err}$ with $N=4$ versus the number of uses $M$ for two types of protocols:
 a Bell biphoton state in both a signal-idler (red) and an idler-free (black) setup. The detection error probability of the target channel with zero acceleration,
among reference channels with
(a) low acceleration ($q_{1}=0.1$), and (b) high acceleration ($q_{1}=0.5$).
}\label{fig3}
\end{figure}

\begin{figure}[ht]
\centering
\includegraphics[width=0.49\textwidth]{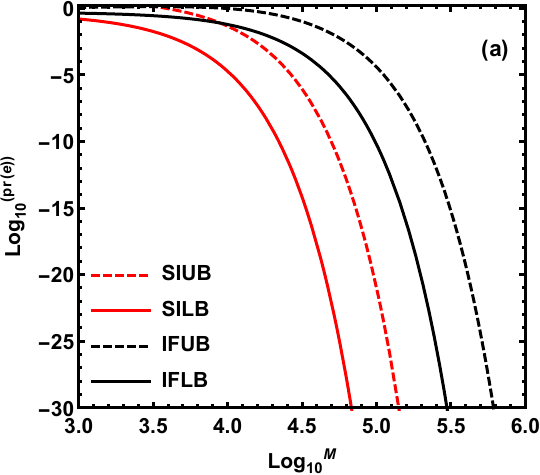}
\includegraphics[width=0.49\textwidth]{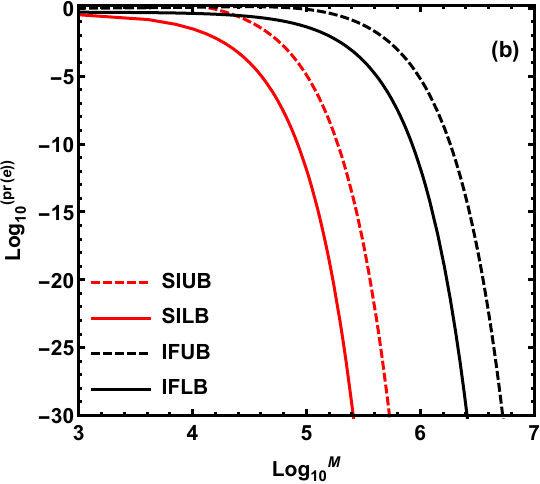}
\caption{(Color online) Quantum channel-position finding error probability $P^{N,M}_{err}$ with $N=4$  versus the number of uses $M$ for two types of protocols:
 a Bell biphoton state in both a signal-idler (red) and an idler-free (black) setup. The detection error probability
of the target channel with nonzero
acceleration ($q_{2}=q_{1}+0.01$),
among reference channels with
(a) low acceleration ($q_{1}=0.1$), and (b) high acceleration ($q_{1}=0.5$).
}\label{fig4}
\end{figure}

If the target channel is a zero acceleration channel within a series of Unruh reference channels, the probe scheme performance is illustrated in Fig. (\ref{fig3}).
It can be seen that for a given $M$, the error probability of $p_{\mathrm{err}}^{\mathrm{SI,U}}$ and $p_{\mathrm{err}}^{\mathrm{SI,L}}$ are both lower than $p_{\mathrm{err}}^{\mathrm{IF,L}}$. This result shows that the error probabilities bound in the SI protocol offer robust advantages, including both the MGA and MPA, along with a scaling advantage in the error exponent.
Figs. (\ref{fig3}a) and (\ref{fig3}b) show that this conclusion remains valid regardless of whether the reference channel is cryogenic or high-temperature. We also find that the MGA function, which guarantees the advantage of the SI, increases as the number of copies of transmitted modes $M$ increases.

If the single target channel is subjected to a nonzero acceleration $q_{2}=q_{1}+0.01$, the temperatures of the target channel and the reference channel become very close, which makes it difficult to distinguish.
Fig. (\ref{fig4}) illustrates the detection error probabilities versus the number of modes $M$
for the SI and IF protocols.
 When resolving two Unruh temperature channels, the SI scheme requires a larger number of copies $M$ compared to the IF scheme, which exhibits both MPA and MGA. We can conclude that the IF protocols for channel localization are only feasible with a very low number of copy probes setting, in which $p_{\mathrm{err}}^{\mathrm{SI,U}}$ is greater than $p_{\mathrm{err}}^{\mathrm{IF,L}}$.
 This demonstrates that a lower probability of detection error can be achieved by increasing the number of copy states.
Comparing Fig. (\ref{fig4}a) and Fig. (\ref{fig4}b), we observe that both schemes perform well in locating a range of cryogenic reference channels, given that high temperatures tend to attenuate the initial quantum correlation.

Based on the analysis, we observe that the SI scheme exhibits significant advantages in the CPF task of Unruh channels, particularly when employing a large number of copy states.
Upon performing the calculations, it becomes evident that the residual quantum correlation of the final state in the SI scheme surpasses that of the IF scheme, enhancing its efficiency in distinguishing between different channels.
In other words, the similarity between the signal and idler states reduces the error probability of the CPF task.
 However, it is worth noting that the IF protocol eliminates the need for idler assistance to achieve quantum advantages in some of the most relevant discrimination scenarios, thereby relaxing practical requirements for prominent quantum sensing applications.

\section{conclusion}\label{conclusion}

This paper has presented performance comparisons among various channel position schemes for the CPF problem, considering channels with different temperatures. Our detection scheme is conducted under energy constraints, specifically utilizing an average of only one photon per channel for detection of the entire channel array.
We consider the geometric characterization of the Unruh channel in Appendix A, where each channel is represented as an operator resulting from the interaction of an UD with its environment. We investigate the task of determining the location of two or more given quantum channels by exploring the SI and IF protocols. The objective is to pinpoint the position of a target Unruh channel within a sequence of reference channels.

 In the task of locating between two Unruh channels, we calculated the output fidelity of CPF to test multiple quantum hypotheses. This provides upper and lower bounds on the error probability, even in cases of small differences in channel temperature and the number of probe states is insufficient.
 When performing the task of distinguishing the zero-temperature channel and the Unruh channel, we stressed out that the SI strategy outperforms the IF strategy, exhibiting both MGA and MPA across entire value regions.
 When resolving the target channel and reference channel with very close temperatures, the IF scheme is effective only in a scenario with a very low number of copy probes.
These findings not only demonstrate that augmenting the number of copy probes can exponentially enhance the efficiency of the detection strategy, but also offer a theoretical framework for laboratories to employ diverse detection protocols for detection tasks.
 We hope that our results will stimulate further research on the discrimination of quantum operations. Our main lesson is that the residual feeble quantum correlation may offer an enormous performance advantage despite its being used in an entanglement-breaking scenario.

 % Next, we introduce the signal-idle protocol and  idle-free protocol to perform the CPF of Unruh channel.
 %We are therefore interested in experimentally determining which of these two situations, if any, occurs in practice. Under the assumption that one of these two alternatives is correct it is possible to frame the problem of determining which of the two is correct in terms of a binary hypothesis test.

\begin{acknowledgments}
This work is supported by the National Natural Science Foundation of China under Grant No. 12122504, No.12203009, No.12374408 and No. 12035005; the innovative research group of Hunan Province under Grant No. 2024JJ1006; the Natural Science Foundation of Hunan Province under grant No. 2023JJ30384; and the Hunan Provincial Major Sci-Tech Program under grant No.2023ZJ1010.
\end{acknowledgments}

%%%%%%%%%%%%%%%%%%%%%%%%%%%%%%%%%%%%%%%%%%%%%%%%%%%%%%%%%%%%%%%%%%%%%%%%%%%%%%%%%%%%%%%%%%%%%%%%%%%%%%%%%%%%%%%%%%%%%%%%%%
\appendix{\bf {Appendix A: The characterization of Unruh channel}}

In this appendix, we discuss the channel characterization of Unruh radiation on the accelerated detectors.
 Consider a two-level semi-classical UD detector \cite{UD2,UD3},
in which the detector follows a classical world line, while its degrees of freedom are quantum.
For a two-qubit system involving Alice and Rob, the detectors carried by Alice remain static, while Rob's detector undergoes uniform acceleration $a$ along the $x$-direction, and its duration of motion is denoted as $\Delta$.
The world line of Rob is described as
\begin{equation}
   t(\tau ) = a^{-1}\sinh{a\tau},\,  x(\tau ) = a^{-1}\cosh{a\tau}, \,  y(\tau) =z (\tau) =0,
\end{equation}
where $\tau$ is the detector's proper time.

The initial state of the total system (detector+field) is given by
\begin{equation}
|\Psi^{AR \phi}_{-\infty} \rangle = |\Psi^{AR}_{-\infty}\rangle \otimes |0_M\rangle,
\label{initialstate}
\end{equation}
where $|\Psi^{AR}_{-\infty}\rangle = \alpha|0_{A}\rangle |1_{R}\rangle+\beta|1_{A}\rangle|0_{R}\rangle$
denotes the initial state shared by Alice's (A) and Rob's (R) detectors, with $|\alpha|^{2}+|\beta|^{2}=1$.
$|0_{M}\rangle$ represents that the external scalar field is in Minkowski vacuum.

The total system Hamiltonian can be expressed as
\begin{equation}
H= H_A +H_R + H_{KG} + H^{R\phi}_{\rm int},
\end{equation}
where $H_{A}=\Omega A^{\dag}A$, $H_{R}=\Omega R^{\dag}R$, and $\Omega$ represents
the detectors' energy gap.
$H_{KG}$ represents the Hamiltonian for the free  Klein-Gordon field.
The accelerated detector Rob is coupled to a scalar field $\phi(x)$ through the
interaction Hamiltonian \cite{UD1}
\begin{equation}\label{interH}
H^{R\phi}_{\rm int}(t)=
\epsilon(t) \int_{\Sigma_t} d^3 {\bf x} \sqrt{-g} \phi(x) [\psi({\bf x})R +
                           \psi^{*}({\bf x})R^{\dagger}],
\end{equation}
where $g\equiv {\rm det} (g_{ab}) $ and ${\bf x}$ represent the coordinates defined on the Cauchy surface $\Sigma_{t = {\rm const}}$ associated with a suitable time-like isometry. The smooth function $\epsilon$ describes the coupling action between the detector and field, enabling the detector to remain active for a proper time interval $\Delta$. $\psi({\bf x})$  is a smooth compact support complex-valued function indicating that the detector only interacts with the field in the neighborhood of its world line.
\iffalse

In the interaction picture the evolved state $|\Psi^{D \phi}_{t}\rangle$ will then be given by
\begin{equation}
|\Psi^{D \phi}_t \rangle =
T \exp\left[-i\int_{-\infty}^t dt' H_{\rm int} ^I(t')\right] |\Psi^{D \phi}_{-\infty} \rangle,
\label{Dyson1}
\end{equation}
where $T$ is the time-ordering operator,
while $H_{\rm int}^I (t) = U^{\dagger}_0(t) H_{\rm int} (t) U_0 (t)$,
with $U_0 (t)=\mathrm{exp}(-iH_0 (t))$.
 By using Eq.~(\ref{interH}), we can rewrite
\begin{equation}
|\Psi^{D \phi}_{t} \rangle =
T \exp \left[-i\int d^4x\sqrt{-g}\phi(x) (fD + \overline{f}D^{\dagger})\right] |\Psi^{D \phi}_{-\infty} \rangle,
\label{Dyson2}
\end{equation}
where
$f \equiv \epsilon(t) e^{-i\Omega t}\psi ({\bf x})$
is a compact support complex function defined in Minkowski spacetime.
\fi

In the interaction picture, we consider the first-order perturbation, and the final state $|\Psi^{AR \phi}_{t}\rangle$ is
\begin{equation}
|\Psi^{AR \phi}_{t} \rangle
= [I - i(\phi(f)R + \phi(f)^{\dagger} R^{\dagger}) ] |\Psi^{AR \phi}_{-\infty} \rangle,
\label{primeira_ordem}
\end{equation}
where
\begin{eqnarray}
\phi(f) & = & i [a_{RI}(\bar{\lambda})-a_{RI}^{\dagger}(\lambda)]
\label{phi(f)}.
\end{eqnarray}
\iffalse
{\color{blue}is an operator valued distribution obtained by smearing out the field operator
by the testing function $f$ above.}
 $a(\overline{u})$ and $a^{\dagger} (u)$ are annihilation and creation
operators of $u$ mode, the operator $K$ takes the positive-frequency part of the solutions of Klein-Gordon equation with respect to the timelike isometry, and
\begin{equation}
Ef = \int d^4x'\sqrt{-g(x')} [G^{\rm adv}(x, x')-G^{\rm ret}(x, x')] f(x'),
\label{Ef}
\end{equation}
where
$G^{\rm adv}$ and $ G^{\rm ret}$ are the advanced and retarded Green functions, respectively.
Next, by imposing that $\epsilon(t)$ is a very slow-varying function of time
compared to the frequency $\Omega$ and that $\Delta \gg \Omega^{-1}$,
then $f$ is an approximately positive-frequency function, i.e.,
$KEf\approx Ef$ and $KE\overline{f}\approx 0$ \cite{coupling1}. Now, by defining
$\lambda \equiv -KEf$,
we take $\phi(f)\approx i a^{\dagger}(\lambda)$
and Eq.~(\ref{primeira_ordem}) as
\fi
Here $a_{RI}(\overline{\lambda})$ and $a_{RI}^{\dagger} (\lambda)$ represent annihilation and creation operators for the $\lambda$ mode.
\iffalse
we obtain Eq.~(\ref{primeira_ordem}) as
\begin{equation}
|\Psi^{AR \phi}_{t} \rangle=
(I + a_{RI}^{\dagger}(\lambda)R -  a_{RI}(\overline{\lambda})R^{\dagger} ) |\Psi^{AR \phi}_{-\infty} \rangle.
\label{primeira_ordem_2}
\end{equation}

The evolution of the whole system follows the Schr$\mathrm{\ddot{o}}$dinger equation, so the first-order expansion of the evolution equation of the system consisting of the detector and the scalar field in the interaction picture \cite{UD1}
\begin{equation}
|\Psi^{D \phi}_{\infty} \rangle=
(I + a^{\dagger}(\lambda)D -  a(\overline{\lambda})D^{\dagger} ) |\Psi^{D \phi}_{-\infty} \rangle.
\label{first-order}
\end{equation}
where $|\Psi^{D \phi}_{-\infty} \rangle$ is the initial state. $\lambda \equiv -KEf$, where $f \equiv \epsilon(t) e^{-i\Omega t}\psi ({\bf x})$,
$K$ denotes the operator for the positive frequency solution of the Klein-Gordon equation, and $E$ is
 the difference between the advance and retard Green function.
\fi
 By inserting Eq. (\ref{phi(f)}) into Eq. (\ref{primeira_ordem}), we obtain
 \begin{eqnarray}
| \Psi^{AR \phi}_{t}\rangle
& = &
|\Psi^{AR \phi}_{-\infty} \rangle
 + \alpha |0_{A}\rangle|0_{R}\rangle
 \otimes(a_{R I}^{\dagger}(\lambda)|0_M\rangle)
 +  \beta|1_{A}\rangle|1_{R}\rangle \otimes(a_{R I}(\overline{\lambda})|0_M\rangle).
\label{evolutionAUX}
\end{eqnarray}
The Bogoliubov transformations between the Rindler operators and the operators annihilating the Minkowski vacuum can be expressed as follows \cite{UD2,UD3}:
\begin{equation}
\begin{aligned}
a_{R I}(\bar{\lambda}) & =\frac{a_M\left(\overline{F_{1 \Omega}}\right)+e^{-\pi \Omega / a} a_M^{\dagger}\left(F_{2 \Omega}\right)}{\left(1-e^{-2 \pi \Omega / a}\right)^{1 / 2}}, \\
a_{R I}^{\dagger}(\lambda) & =\frac{a_M^{\dagger}\left(F_{1 \Omega}\right)+e^{-\pi \Omega / a} a_M\left(\overline{F_{2 \Omega}}\right)}{\left(1-e^{-2 \pi \Omega / a}\right)^{1 / 2}},
\end{aligned}
\end{equation}
where
$F_{1 \Omega}=
\frac{\lambda+ e^{-\pi\Omega/a}\lambda\circ w}{(1- e^{-2\pi\Omega/a})^{{1}/{2}}}$,
$F_{2 \Omega}=
\frac{\overline{\lambda\circ w}+ e^{-\pi\Omega/a}\overline{\lambda}}{(1- e^{-2\pi\Omega/a})^{{1}/{2}}}$, and
$w(t, x)=(-t, -x)$ represents the wedge reflection isometry.

Then we obtain the reduced density matrix of the detector's state by tracing out the degrees of freedom of the external field
\begin{equation}
\begin{aligned}
\rho_{t}^{AR} & =\left\|\Psi_{t}^{AR \phi}\right\|^{-2} \operatorname{Tr}_\phi\left|\Psi_{t}^{AR \phi}\right\rangle\left\langle\Psi_{t}^{AR \phi}\right| \\
& =
\left(	\begin{array}{cccc}
\mathcal{C} &  0 &  0 & 0\\
		0& |\alpha|^{2}\mathcal{A}  &\alpha\beta \mathcal{A} & 0 \\
		0& \alpha\beta\mathcal{A} &   |\beta|^{2}\mathcal{A}  & 0 \\
0&  0&  0& \mathcal{B}
	\end{array}\right),
\end{aligned}
\end{equation}
where
\begin{equation}
\begin{aligned}
\mathcal{A} & =\frac{1-q}{\left(1-q\right)+ \nu^2\left(|\alpha|^{2}+|\beta|^{2}q\right)}, \\
\mathcal{B} & =\frac{\nu^2|\beta|^{2}q}{\left(1-q\right)+ \nu^2\left(|\alpha|^{2}+|\beta|^{2}q\right)}, \\
\mathcal{C} & =\frac{\nu^2|\alpha|^{2}}{\left(1-q\right)+ \nu^2\left(|\alpha|^{2}+|\beta|^{2}q\right)},
\end{aligned}
\end{equation}
with the parametrized acceleration $q \equiv e^{-2 \pi \Omega / a}$. The effective coupling between the detector and the scalar field is
$\nu^2 \equiv\|\lambda\|^2=\frac{\epsilon^2 \Omega \Delta}{2 \pi} e^{-\Omega^2 \kappa^2}$ \cite{curved1,coupling3,UD2,UD3},
where $\Omega^{-1}\ll\Delta$ is necessary for the validity of the above definition.
In the present work the  coupling parameter is constrained to $\nu^2\rightarrow0$ to ensure the validity of the perturbative approach.
Notably, $q$ is a monotonic function of acceleration $a$, and $q\rightarrow0$ corresponds to zero acceleration. These facts suggest that the Unruh effect can be interpreted as a noisy quantum channel.

The dynamics of open quantum systems can be characterized as follows.
The evolution from the detectors initial state $\rho^{AR}_{-\infty}$ to the final state $\rho^{AR}_{t}$ can alternatively be expressed as:
\begin{equation}\label{evolution}
\rho^{AR}_{t}  =U(t)\rho^{AR}_{-\infty}U^{\dagger}(t),
\end{equation}
where $U(t)$ is the propagator of the joint system dynamics from the initial time to the final time.
The object of interest is the subsystem $R$, whose state at all times $t$ is governed according to the standard quantum mechanical prescription by the following quantum dynamical process:
\begin{equation}\label{evolution}
\rho^{R}_{t}  =\text{Tr}_{B}\left[U(t)\rho^{AR}_{-\infty}U^{\dagger}(t)\right].
\end{equation}
%Each channel is characterized by an accelerated detector which interacts with its surroundings
The quantum dynamical process
can be described by a quantum map denoted as:
\begin{equation}\label{map111}
\mathcal{C}_{\rho}  =\sum^{\infty}_{j}M_{j}\rho_{0} M_{j}^{\dagger},
\end{equation}
where $\rho_{0}$ is an arbitrary initial state, and $M_{j}$ is an operator for different dynamical evolution processes.
Based on the above analysis and calculation,
the operators $M_{\nu}^R$ acting on Rob can be characterized by the following Choi Matrix:
\be
M_1^{R}=\left(
\begin{array}{cclr}
\sqrt{1-q}&0\\
0&\sqrt{1-q}
\end{array}\right),~~
M_2^{R}=\left(
\begin{array}{cclr}
0&0\\
v\sqrt{q}&0
\end{array}\right),~~
M_3^{R}=\left(
\begin{array}{cclr}
0&v\\
0&0
\end{array}\right).
\label{eq:ukraus}
\ee
If we are considering the subsystem $\rho^{A}_{t} $, with $M_{\mu}^A$ represents identical because the detector of Alice remains static and is switched off.

%\newpage


\begin{thebibliography}{99}

\bibitem{Unruh1}
W. G. Unruh, \textit{Phys. Rev. D} {\bf 14}, 870 (1976).

\bibitem{UD1}
W. G. Unruh and R. M. Wald, \textit{Phys. Rev. D} {\bf 29}, 1047 (1984).

\bibitem{HKG}
 L. C. B. Crispino, A. Higuchi and G. E. A. Matsas, \textit{Rev. Mod. Phys.} {\bf 80}, 787 (2008).

 \bibitem{Unruh2}
W. G. Brenna, E. G. Brown, R. B. Mann, and E. Mart\'{i}n-Mart\'{i}nez, \textit{Phys. Rev. D} {\bf 88}, 064031 (2013).


\bibitem{gendeg1}
I. Fuentes and R. B. Mann, \textit{Phys. Rev. Lett.} {\bf 95}, 120404 (2005).

\bibitem{exp3}
E. Mart\'{i}n-Mart\'{i}nez, I. Fuentes, and R. B. Mann,  \textit{Phys. Rev. Lett.} {\bf 107}, 131301 (2011).

\bibitem{gendeg2}
E. Mart\'{i}n-Mart\'{i}nez and J. Le$\acute{o}$n,  \textit{Phys. Rev. A} {\bf 80}, 042318 (2009).

\bibitem{curved1}
R. M. Wald, Quantum Field Theory in Curved Spacetimes and Black Hole Thermodynamics (The University of Chicago Press, Chicago, 1994).

\bibitem{curved2}
J. Wang and J. Jing, \textit{Phys. Rev. A} {\bf 82}, 032324 (2010).

\bibitem{UD2}
A. G. S. Landulfo and G. E. A. Matsas,  \textit{Phys. Rev. A} {\bf 80}, 032315 (2009).

\bibitem{UD3}
L. C. C$\acute{e}$leri, A. G. S. Landulfo, R. M. Serra, and G. E. A. Matsas,  \textit{Phys. Rev. A} {\bf 81}, 062130 (2010).

\bibitem{UD4}
X. Liu, Z.  Tian, J. Wang, and J. Jing, \textit{Phys. Rev. D} 97, 105030 (2018).

\bibitem{coupling3}
J.  Wang, Z.  Tian, J. Jing, and H. Fan,  \textit{Phys. Rev. A} {\bf 93}, 062105 (2016).

\bibitem{HK1}S. W. Hawking, \textit{Nature (London)} {\bf 248}, 30 (1974).

\bibitem{HK2}G. W. Gibbons and S. W. Hawking, \textit{Phys. Rev. D} {\bf 15}, 2738 (1977).

\bibitem{MK}
M. Kalinski, \textit{Laser Phys.} {\bf 15},
1367 (2005), arXiv:quant-ph/0501172.

\bibitem{curved4}
Q. Liu, S.-M. Wu, C. Wen, and J. Wang,  \textit{Sci. China-Phys. Mech. Astron.}
{\bf 66}, 120413 (2023).

\bibitem{curved5}
S.-M. Wu, H.-S. Zeng, T. Liu, \textit{New J. Phys.} {\bf 24}, 073004 (2022).

\bibitem{curved6}
 A. Mukherjee, S. Gangopadhyay, and A. S. Majumdar, \textit{Phys. Rev. D} {\bf 108}, 085018 (2023).

\bibitem{experimental1}
W. G. Unruh, \textit{Phys. Rep.} {\bf 307}, 163 (1998).

\bibitem{experimental2}
R. Sch$\mathrm{\ddot{u}}$tzhold, G. Schaller and D. Habs,  \textit{Phys. Rev. Lett.} {\bf 97}, 121302 (2006).

\bibitem{experimental3}
Z. Tian, J. Wang, J. Jing, A. Dragan, \textit{Ann. Phys.} {\bf377}, 1-9(2017).


\bibitem{experimental4}
E. T. Akhmedov, and K. Gubarev,  arXiv:2310.02866 (2023).


\bibitem{open1}
I. Pe\~{n}a, D. Sudarsky, \textit{Found. Phys.} {\bf44}, 689 (2014).

%\bibitem{QHT1}C. Helstrom, Quantum Detection and Estimation Theory, Mathematics in Science and Engineering: A Series of Monographs and Textbooks (Academic Press, New York, 1976).
%\bibitem{QHT2}J. A. Bergou, \textit{J. Mod. Opt.} {\bf57}, 160 (2010).
\bibitem{channel1}
 J. L. Pereira, Q. Zhuang, and S. Pirandola,
 \textit{Phys. Rev. Res.} {\bf2}, 043189 (2020).

\bibitem{channel2}
Q. Zhuang and S. Pirandola,
 \textit{Phys. Rev. Lett.} {\bf125}, 080505 (2020).

\bibitem{CPF0}
M. F. Sacchi, \textit{Phys. Rev. A} {\bf 72}, 014305 (2005).

\bibitem{CPF1}
Q. Zhuang and S. Pirandola, \textit{Commun. Phys.} {\bf 3}, 103 (2020).


\bibitem{CPF2}
J. L. Pereira, L. Banchi, Q. Zhuang, and S. Pirandola, \textit{Phys. Rev. A} {\bf103}, 042614 (2021).

\bibitem{CPF3}
A. Karsa, J. Carolan and S. Pirandola, \textit{Phys. Rev. A} {\bf105}, 023705 (2022).

\bibitem{DC}
A. Arqand, L. Memarzadeh and S. Mancini,
 \textit{Phys. Rev. A} {\bf102}, 042413 (2020).

\bibitem{ADC}
M. Rexiti and S. Mancini,
 \textit{J. Phys. A: Math. Theor.} {\bf54}, 165303 (2021).

\bibitem{noisy1}
D. Ahn, \textit{Phys. Rev. A} {\bf 98}, 022308 (2018).

\bibitem{SI1}
S. Lloyd, \textit{Science} {\bf 321}, 1463 (2008).

\bibitem{QI1}
S. H. Tan, B. I. Erkmen, V. Giovannetti, S. Guha, S. Lloyd, L. Maccone, S. Pirandola, and J. H. Shapiro, \textit{Phys. Rev. Lett.} {\bf 101}, 253601 (2008).

\bibitem{IF1}
C. Harney and S. Pirandola, \textit{npj Quantum
Information} {\bf 7}, 153 (2021).


\bibitem{symme1}
E. Bagan, J. A. Bergou, S. S. Cottrell, and M. Hillery, \textit{Phys. Rev. Lett.} {\bf 116}, 160406 (2016).

\bibitem{symme2}
D. Qiu and L. Li, \textit{Phys. Rev. A} {\bf 81}, 042329 (2010).

\bibitem{LUB1}
S. Zhang, Y. Feng, X. Sun, and M. Ying, \textit{Phys. Rev. A} {\bf 64}, 062103 (2001).

\bibitem{fidelity1}
R. Jozsa, \textit{J. Mod. Opt.} {\bf 41}, 2315 (1994).

\bibitem{fidelity2}
 C. Harney, L. Banchi, and S. Pirandola,
\textit{Phys. Rev. A} {\bf 103}, 052406 (2021).

\bibitem{QI2}
A. Karsa, G. Spedalieri, Q. Zhuang, and S. Pirandola, \textit{Phys. Rev. Res.} {\bf 2}, 023414 (2020).

\bibitem{QI3}
R. Nair and M. Gu, \textit{Optica} {\bf 7}, 771 (2020).

\bibitem{Spectroscopy1}
H. Shi, Z. Zhang, S. Pirandola, and Q. Zhuang, \textit{Phys. Rev. Lett.} {\bf 125}, 180502 (2020).

\bibitem{QD1}
S. Pirandola, \textit{Phys. Rev. Lett.} {\bf 106}, 090504 (2011).

\bibitem{ops}
H. P. Breuer and F. Petruccione,  \textit{The Theory of Open Quantum Systems} (Oxford University Press, Oxford), (2002).

\end{thebibliography}
\end{document}